# Ultralight Plasmonic Structural Color Paint

Pablo Cencillo-Abad[1,†], Daniel Franklin[1, 2, †], Pamela Mastranzo-Ortega[1], Debashis Chanda[1, 2, 3]

[1] *NanoScience Technology Center, University of Central Florida, 12424 Research Parkway Suite 400, Orlando, Florida 32826, USA.*

[2] *Department of Physics, University of Central Florida, 4111 Libra Drive, Physical Sciences Bldg. 430, Orlando, Florida 32816, USA.*

[3] *CREOL, The College of Optics and Photonics, University of Central Florida, 4304 Scorpius St., Orlando, Florida 32816, USA.*

*[†]Equal contribution.*

*Correspondence and requests for materials should be addressed to D.C. (email: Debashis.Chanda@ucf.edu).*

In nature, vibrant colors as those of many butterflies, birds, octopuses, or fishes, arise from microscopically textured surfaces. These vivid colors result from the coherent interaction between light and the structural arrangement of colorless materials found in their skin. In contrast, all manmade colors are pigment based and rely on the molecular absorption of their constituents, with each color requiring a different molecule. While traditional pigment-based colorants offer a viable commercial platform for large-volume and angle-insensitiveness, they are limited by their resolution, instability in atmosphere, color fading, and severe environmental toxicity. These widely used pigment-based paints are destroying the environment, aquatic life, and adversely affecting global warming by working as heat traps. However, till date all attempts to industrial production of polarization and angle independent full-range structural colors have failed due to the angle-dependent colors and fabrication challenges. Here, we present a subwavelength plasmonic cavity that generates color by the hybridization of a metallic self-assembly with an ultrathin optical cavity. This configuration offers both polarization and angle insensitiveness, while simultaneously providing a full-color gamut of vibrant structural color paints. In this work, we presented this unique structural color generation mechanism and demonstrated color generation of these structures across the entire visible spectrum by tuning just the structural parameters. Further, akin to traditional mixing of different pigments to produce new colors, we demonstrated in a unique way lateral as well as vertical "mixing of structures" to expand the available color palette. The self-assembly facilitates the growth of the structure on large areas and non-conventional substrates in both diffuse and specular coloration modes. Growing the stack on a sacrificial layer

**we produce a self-standing nanostructured color platform that, when mixed with a binder, can be transferred to any surface to impart full coloration with a single sub-micron layer of pigment. This structural color platform offers a highly integrable ultra-lightweight solution that bridges the gap from proof-of-concept to real-world industrial applications of non-toxic, fade resistant, and environmentally friendly colorants.**

## Introduction

Color presents one of the richest sources of sensorial information in our daily lives. Throughout history, the fascination with colors has driven human efforts to produce newer and better colorants. From the Paleolithic cave paintings to the development of the first synthetic dyes in the mid nineteenth century the quest for purer, fade-resistant, and environmentally-friendly colorants has remained very active. In the last decades, adding to purely decorative applications in textile, cosmetics, or food industries, colorant research has found relevance, among others, in display technologies, optical storage, sensing and therapeutics, or functional coatings[1,2].

Color engineering can be achieved by controlling the colorant's absorptive or reflective response to white light. All commercial colorants/pigments are based on absorption mechanisms. These colorants absorb photons of energies overlapping with their molecular electronic transitions. Contrarily, photons with energies not matching these discrete transitions will be reflected and registered as color by an observer. Although chemical colorants can be produced in large amounts, most of them are composed of toxic materials difficult to remove in the recycling process and are responsible for the pollution of our lifeline on earth—water[3]. Being chemically unstable, many colorants fade with time, a process accelerated with higher temperatures or light exposure. Furthermore, as volumes of several microns are needed to obtain enough color saturation, they suffer from low resolution. In contrast, instead of controlling the absorption of light, structural colorants control the way the light is reflected or scattered. Structural color is the result of optical phenomena produced by micron- and nano-scale structures[4]. Remarkably, when in bulk, the material constituents show completely different hues or are even colorless. Colors generated by engineered structures such as photonic crystals[5–9] or metasurfaces[10–13], have received increasing attention in recent years for their striking advantages over chemical colorants. Characterized by their intense brilliance and saturation, they exhibit larger stability to chemical reagents, harsh environmental conditions, and high illuminating intensities[14,15]. Additionally, they can offer dynamic tunability and resolutions beating the diffraction limit, both essential for display applications[16–18]. Due to the geometrical nature of their response, however, structural colors usually present directional effects, i.e. their color varies with the positioning of the observer and the angle and polarization of the incident light. More importantly, many proposed architectures rely on the use of costly and low-throughput nanofabrication techniques not compatible with mass-production. Overall, these constrains prohibit the commercial viability of all previously reported structural color technologies. It is therefore not surprising that, to date, no angle-independent structural paints are available in the marketplace[19].

Here, we present a subwavelength plasmonic cavity that overcomes these challenges while offering a tailorable platform for rendering angle and polarization independent vivid structural colors by coupling incident light with gap-plasmons. The structures are fabricated through a large-area, highly versatile, and reproducible technique where aluminum nanoislands are self-assembled in an electron beam evaporator on top of a transparent oxide-coated aluminum mirror. The optical response of these artificially engineered nanostructures can be spectrally tuned across the entire visible spectrum to form a full color gamut by controlling the gap-plasmon dispersion via the geometrical parameters. In the proposed architecture the subwavelength optical cavity ensures a large degree of angle insensitivity while the stochastic nature of the self-assembled layer results in polarization independence and near 100% absorption at selected spectral bands. The evaporation process, relying only on widespread industrial techniques, is compatible with many substrates, and takes on their scattering properties to render diffuse and specular coloration modes when utilizing micro-corrugated or flat surfaces, respectively. E-beam evaporators are widely employed in industries such as electronics, semiconductors, optics, and even aerospace, to name a few. Moreover, Lexus Blue, the only industrially produced simple Fabry-Perot resonance based structural color is actually fabricated with ebeam evaporators[20] We present mechanisms for expanding the available color space through lateral and vertical mixing of structures, similar to traditional pigment mixing schemes. Finally, to demonstrate the commercial capabilities of our platform for inorganic metallic structural coloration, we formed bi-directional structures on a water-soluble sacrificial layer that resulted in omni-directional color flakes. These structural color flakes were then mixed with a commercial binder to develop self-standing structural color paints hundreds of times lighter than commercially available paints[21]. Conventional chemical coloration relies on volumetric absorption of light to produce a color. In contrast to the several microns required for commercial paints, our ultrathin paint can impart full coloration with a thickness of only 150 nm. Consequently, this huge lateral area (few 10s of μm) to thickness (100 – 150 nm) ratio makes it the lightest paint in the world with a surface density of only 0.4 g/m$^2$. For comparison, while a Boeing 747 requires 500 kg of paint[22], our ultralight paint would require about 1.3 kg, an astonishing potential about 400-fold reduction in weight, *SI Appendix I*. Our approach presents the first environmental-friendly, large-scale, multi-color, and self-standing platform for imparting nanostructured coloration to any surface, thus bridging the gap from proof of concept to industrial production.

# Results

**Self-Assembled Plasmonic Surface.** Nature presents a rich variety of both chemical and structural coloration. For example, the pink tint of Formosa azaleas, Figure 1a, is due to the absorption of cyaniding molecules, a type of anthocyanin pigment[23]. In contrast, the bright metallic blue displayed by the Peruvian Morpho didius, Figure 1b, is primarily the result of the way the blue components are scattered by the lameallae nanostructures found in this butterfly's wings[24]. Oftentimes, however, structural color in animals results from the combination of the diffraction and scattering of the outer skin layers, and the molecular absorption of the complementary color by intrinsic pigments of the skin[25]. This critical observation inspired us to produce an absorptive structural pigment where the selective absorption of specific frequencies is the result of the tailored structural resonant response of metallic nanostructures coupled to a subwavelength optical cavity. Specifically, the proposed architecture consists of a highly-packed monolayer of self-assembled aluminum nanoislands on a thin aluminum oxide film that spaces them from the aluminum back-mirror, Figure 1c. In this configuration the aluminum nanoislands resonantly absorb specific wavelengths, while the back mirror strongly back-reflects the non-resonant ones, rendering vivid colors based on colorless materials.

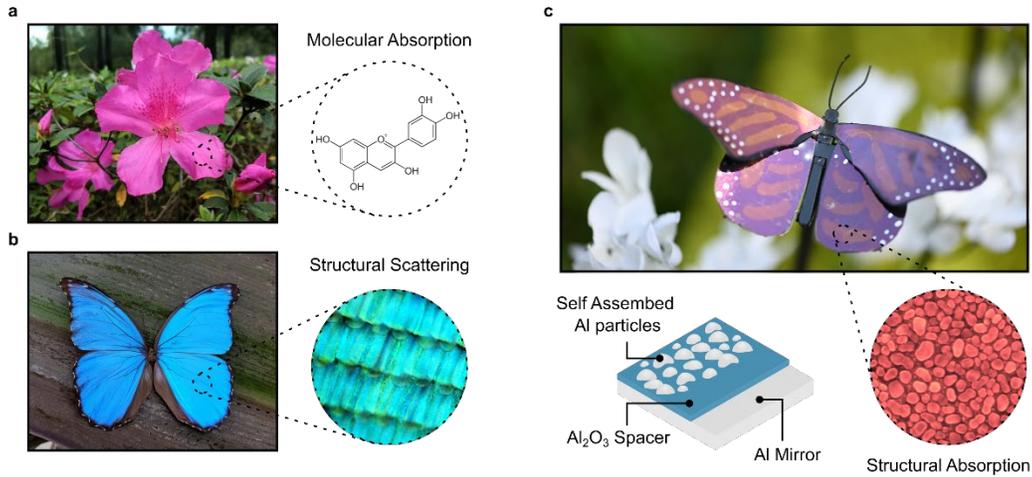

**Figure 1 | Structural Absorption for Color Generation**. **a**, Many chemical substances produce color by selectively absorbing frequencies matching their molecular electronic transitions. Pink color in Formosa azaleas is due to the absorption of cyaniding molecules. **b**, An example of structural coloration is found in the Peruvian Morpho didius. Lamellae nanostructures found in its wings scatter the blue components of incident light generating its characteristic metallic blue. **c**, A subwavelength plasmonic cavity formed by a self-assembly of metallic nanoislands on top of an oxide-coated mirror, generates color by selectively absorbing certain wavelengths and strongly back-reflecting other.

Contrary to other artificial structural schemes that rely on the use of low-throughput, multi-step, top-down techniques such as electron beam lithography or focused ion beam, incompatible with mass-production, the proposed architecture is the result of a naturally occurring nucleation process in an electron beam evaporator. In the self-assembly growth, small clusters of aluminium nanoparticles are formed due to the larger affinity of the aluminium atoms to their own kind over the oxide substrate. With a low enough rate, the evaporation of nanometric films results in a nanoparticles' monolayer that exhibit optical plasmonic resonances. Crucially, this pressure- and temperature-controlled process ensures high reproducibility over broad areas in a single step, lowering the cost of production and enabling large-scale fabrication. The dynamics of the self-assembly process are presented in detail in *SI Appendix A*, while the technical parameters can be found in Methods.

**Optical Response of the Near-Field Coupled Gap Plasmons**. The color produced in the nanostructure is the result of the hybridization of the absorptive response of the aluminum self-assembled monolayer, and the subwavelength cavity formed by this top layer, the aluminum back-mirror, and the dielectric spacer sandwiched in between. Geometrical changes in any of the layers will then result in a change in the perceived color. When ambient light reaches the monolayer, the electric field of the light at select wavelengths can drive the free electrons of the aluminum to oscillate resonantly within the nanoparticles' geometry. This collective oscillation, termed localized surface plasmon resonance, is further affected by the coupling between closely-packed neighboring particles and the presence of the back-mirror interface at a subwavelength distance from the particles' layer. This complex hybridization mechanism results in a gap-plasmon mode that leads to strong optical absorption and tight confinement of the light at the metal/dielectric boundary of the metallic particles at resonant frequencies[26]. The spectral position of the absorption band, and thus the perceived color, depends distinctly on the gap-plasmon dispersion which is hence controlled by three parameters: (1) the size and spatial distribution of the nanoislands, (2) the refractive index of their environment, and (3) the thickness of the spacing layer.

The size of the nanoislands can be simply controlled by tuning the amount of aluminum evaporated. To investigate the range of colors available with this cumulative process we use a shutter that controls the partial exposure of the sample during the evaporation. In this manner, by rotating the sample, we can produce a polar gradient of thicknesses from 0.5 nm to 16 nm, in thickness increments of 0.5 nm corresponding to wedges of approximately 11°, Figure 2b. As the thickness mass is increased neighboring nanoislands coalesce to form larger particles, Figure 2a-top. This increase in the nanoislands' size red-shifts the absorption band and results in different hues and saturations that produce a color palette that covers from the white of the back mirror, at very low thicknesses, to the yellow, magenta, and blue. It should be noted that, being a subtractive color scheme, a red-shift of the absorption band results in a blue-shift in perceived color, as the intensity of blue components in the reflected light augment at the expense of the yellow and red ones. If the process is carried on for long enough, adjacent nuclei can coalesce to form semi-continuous films and, eventually, continuous films, Figure 2a-bottom. The thickness at which the transition from isolated islands to continuous film occurs is the percolation threshold. At thicknesses above the percolation threshold the free electrons of the metal can find paths to move through the self-assembly, eliminating the geometrical confinement necessary for the resonant plasmonic absorption and thus disabling the color production. This can be observed in the gradient wheel sample at higher thicknesses where the blue fades to white, Figure 2b. Further details on the growth dynamics can be found in *SI Appendix A*.

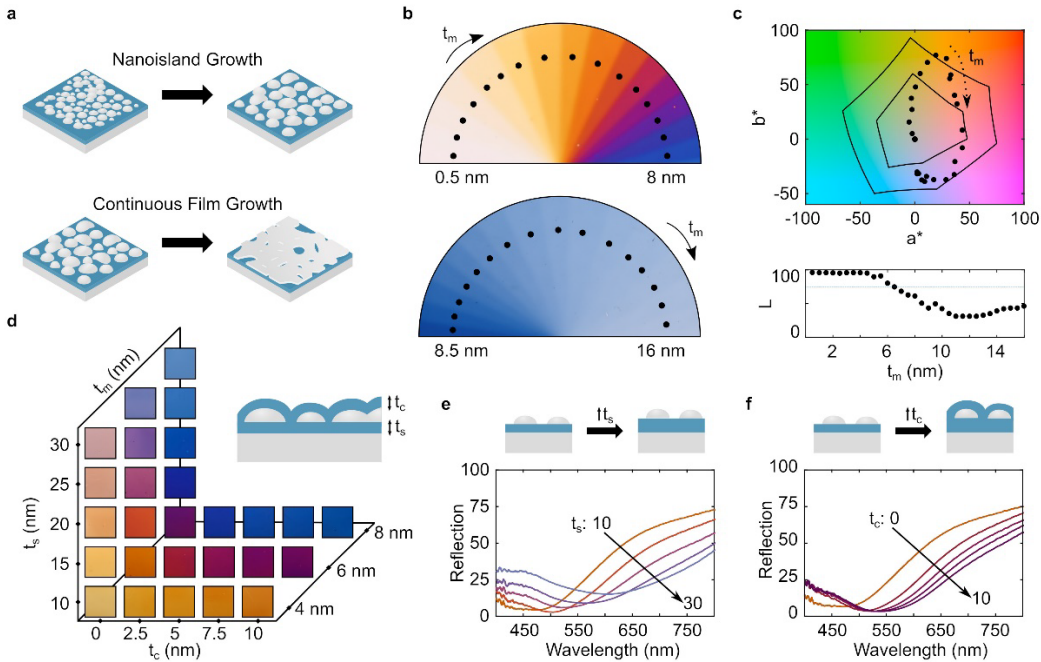

**Figure 2 | Color Space based on Tunable Gap Plasmon Dispersion. a**, In the Volmer-Weber growth mode, the size of the nanoislands can be controlled by tuning the amount of aluminum evaporated –top-. If the process is carried on for long enough semicontinuos films are formed that disable the plasmonic resonances and thus the color –bottom-. **b**, Color polar gradient for thicknesses from 0.5 to 16 nm, for fixed 10 nm spacer. **c**, CIELAB coordinates for the points in the color wheel compared to ISO DIS 15339-2 cold-set newsprint and coated premium paper standards (inner and outer hexagon). **d**, Tuning of the spacer and capping layer thicknesses expands the available color space. **e**, **f**, Show the red-shift of the absorption resonance as the spacer and capping layer thicknesses are increased.

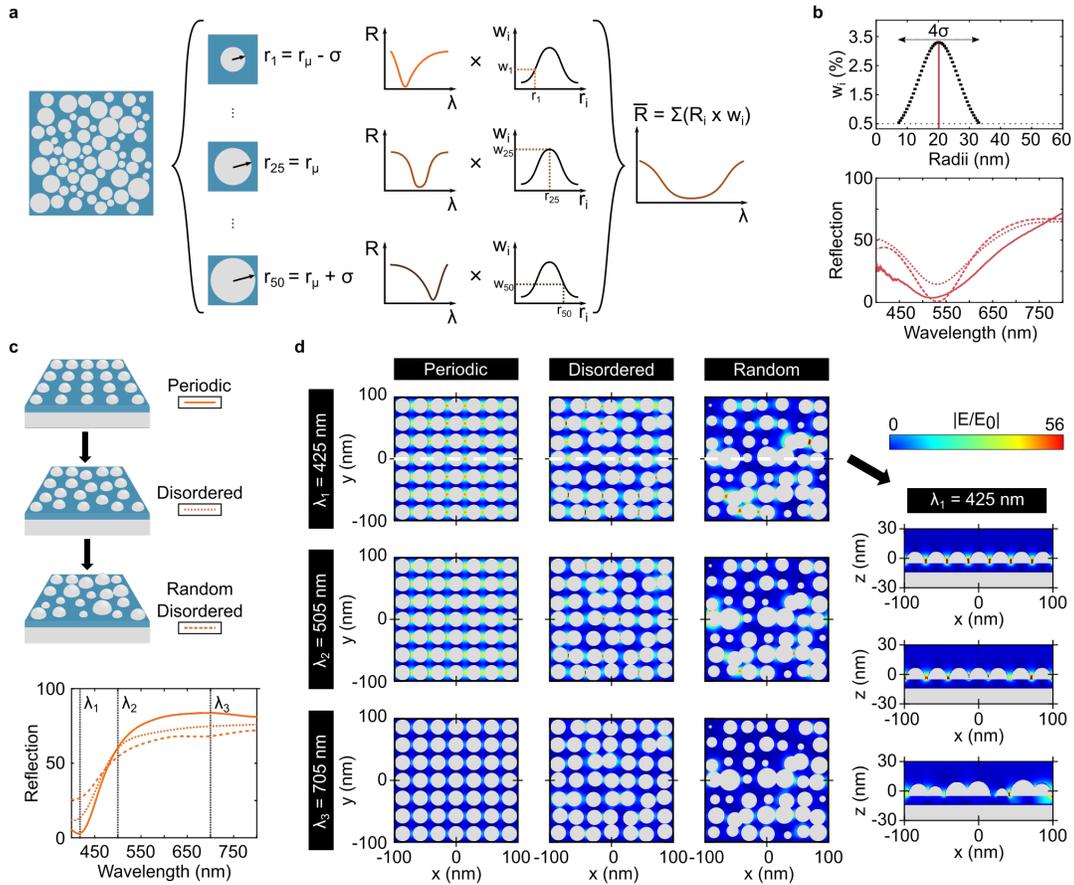

**Figure 3 | Morphology Effect on Optical Response. a**, The inhomogeneous broadening of the optical resonance can be accounted for by introducing size and spatial variability. We simulate the broadening by averaging the reflection curves of 50 particles with radii within 4 standard deviation of the mean value obtained from SEM analysis. **b**, Statistical radii distribution -top- and reflection curves -bottom- for experimental (solid), FDTD mean value (dashed), and weight averaged (dotted). **c**, Reflection curves for FDTD simulations corresponding to 7x7 hemispherical particles with equal size in periodic and disorder arrangement, and random size and disordered arrangement, equivalent to the 5 nm self-assembly. **d**, Electric profiles in three different spectral positions as labeled in panel c.

Together with the spectral shift, the increase in the thickness mass results in a broadening of the optical resonances. We attribute this phenomenon to the inhomogeneous broadening of the nanoparticles' resonances arisen from the doubly random nature of both the morphology and spatial distribution, as can be seen in Figure 3. On the one hand, as thickness mass increases, larger variability of island size can be observed, *SI Appendix Figure 1a*. To further assess this effect, we build a semi-analytical model that defines the total reflection of the monolayer by weight averaging the reflection of periodic islands of 50 hemispherical radii within 4 standard deviation of the mean value as obtained from the SEM analysis for the 8 nm thickness mass, Figure 3a and *SI Appendix B*. This larger morphological variability translates into a reduction in the reflection contrast and saturation of the colors produced, with a distinct broadening of the resonance, Figure 3b. On the other hand, the effect of the spatial distribution can be explained by the well-known dependency of relative position of interacting plasmonic resonators[27]. To evaluate this latter effect, we run a set of simulations for 7x7 hemispherical nanoparticles, for equivalent thickness mass of 4 nm on top of a 10 nm oxide spacer, in periodic square array, disordered array, and disordered randomized sizes, Figure 3c. The reflection curves show additionally a spectral shift that we associate with the different energies of the new available modes resulting from the laterally-hybridization of nanoparticles, modes otherwise forbidden in the symmetric arrangement, *SI Appendix C*. These assumptions are indeed confirmed from the comparison of the electric profiles in-resonance where we observe the strongly confined fields characteristic of the gap plasmon modes, for both ordered and disordered arrangements, as seen in Figure 3d. Interestingly, we observe clearly that while for the ordered structure the dipolar resonance is only excited at in-resonance wavelength, both disordered and random disordered configurations show excitations even well outside the in-resonance spectral position. The inhomogeneous broadening observed is also in good agreement with the expected behavior predicted by the classical formula for dipole-dipole interaction energy given by[28]:

$$W = k_e k_\alpha \frac{|\boldsymbol{p}_1||\boldsymbol{p}_2|}{n_e^2|\boldsymbol{r}_{12}|^3} \qquad (1)$$

where $k_e$ is the Coulomb constant, $k_\alpha$ the orientation factor, $n_e$ the refractive index of the environment, $|\boldsymbol{p}_1|$ and $|\boldsymbol{p}_2|$ the modulus of the dipole moments for two interacting particles, and $|\boldsymbol{r}_{12}|$ the modulus of the distance between them. In this near-field approximation, considering two neighbor particles interacting, if their sizes, and also shapes, show a large variability, it is expected that the dipole modes corresponding to a given illuminating wavelength will be indeed weakly excited, and consequently lower absorption will result with a poorer reflection contrast. Furthermore, the spatial disorder broadening can be understood by averaging the distance between particles, where some of them will be constructively interfering, while others will be out of phase and thus destructively interfering. Finally, it should be noted that unlike transmissive colors, emitting out of a source, all subtractive colors lack purity (paper print vs. LED displays). In this case the high-density packing of the self-assembly plays a critical role in bringing the hybridized modes to the visible range, while ensuring vivid coloration in a single nanometric layer, it exacerbates the resonance broadening resulting from the dynamic depolarization of non-spherical particles, see *SI Appendix D*. Although, all factors considered, spectrally purer colors could be achieved with pre-treatment steps prior to the self-assembly growth, this would be achieved at the expense of the fabrication simplicity offered in our approach. To better understand the coupled mechanism, we also develop a theoretical model, and compared it with FDTD simulations, results of this can be found in *SI Appendix D*.

To assess the quality of the color gamut generated by our self-assembled plasmonic structure, we calculated the L*a*b* coordinates from the reflection spectra of each thickness in the gradient sample (see Methods), and plot them as black dots in the CIELAB color space, Figure 2c. To compare with two color quality standards used in the printing industry, we overlay the standards for the cold-set newsprint and coated premium paper technologies as defined in ISO DIS 15339-2 (inner and outer hexagon respectively). For a substantial portion of the color space, we find that the self-assembled plasmonic color exceeds the newsprint standard, and even matches the quality of some colors as produced in coated premium paper. However, although the color space of the plasmonic structure can be expanded in some regions by careful selection of the other geometrical parameters, due to its subtractive nature, the production of green is prohibited for a single particle layer. To address this limitation, in sections below, we introduce two different color mixing schemes.

Due to the strong field confinement in the metal-dielectric interface plasmonic resonances are extremely sensitive to changes in the environment[28,29]. The addition of a capping layer on top of the self-assembly presents an opportunity to further tune the color response by shifting the resonant spectral position of the nanoislands' layer. For samples corresponding to 4, 6, and 8 nm mass thicknesses, and fixed 10 nm-thick oxide spacer, we monitor the color change as we grow capping layers of alumina in 2.5 nm increments, Figure 2d. Reflection curves for the 6 nm samples can be seen in Figure 2f. Clearly, the presence of the capping layer red-shifts the plasmonic resonance producing colors with higher blue components. This behavior is captured by the classical formula for the dipole-dipole energy interaction presented in *SI Appendix D*. As the thickness of the capping layer increases, more energy is contained within the higher dielectric media and the particle-particle interaction weakens. This causes lower hybridization energies and results in higher resonant wavelengths. The effect of this top layer is of particular importance from the applications point of view. Although aluminum, due to its native oxide layer, is very chemically stable in atmosphere, we found the structures to be fragile to harsh contaminants and physical contact. To address this, we capped samples with a commercial polyurethane clear coat (DuraClear Varnish, Americana). Interestingly, these samples still maintained vivid colors while offering protection to physical contact and larger chemical resistance to spills as can be observed in *SI Appendix E*.

The final element that controls the optical response of the structure is the spacer defined by the thickness of the transparent aluminum oxide spacer layer. As shown in Figure 2d for samples corresponding to 4, 6, and 8 nm mass thickness and varying spacers from 10 to 30 nm, changes in the spacer thickness result in pronounced color changes in the structure. For the 6 nm nanoparticles' layer the reflection curves are shown in Figure 2e. We observe that as the spacer is increased the resonance is shifted to longer wavelengths and the overall reflection levels increase producing less saturated colors. We explain the behavior of the multilayer stack using interference theory of a non-symmetric subwavelength cavity, where the bottom mirror and the top nanostructured self-assembly form the two limiting interfaces, and the ultra-thin dielectric (alumina) spacer sandwiched between them which controls the vertical coupling between two metallic layers. This configuration is essential to achieve the almost-100% levels of absorption in the nanostructured plasmonic self-assembled layer, which occur only when field-enhancement occurs at the nanoparticles layer for wavelengths that fulfill the phase matching condition[30]. In contrast to conventional Fabry-Perot resonators, where the phase is simply accumulated through the propagation in the dielectric and the resonant condition can only be fulfilled for cavity lengths proportional to the wavelength of light, the dispersive nature of the gap-plasmon mode excited on the self-assembled Al nanoislands introduces an interface with non-trivial phase shifts and high losses that can produce absorption resonances even for deeply subwavelength thicknesses well below the resonant wavelength. As the thickness of the spacer is further increased the mismatch between phases results in weaker absorption response and renders less saturated colors. The different nature of this near-field coupled gap-plasmon mode compared to a far-field Fabry-Perot mode, is further verified for large enough spacer thicknesses. When the dielectric spacing layer takes values large enough (multiples of $\lambda/4n_s$), far field effects become dominant and the phase accumulated through propagation can fulfill the resonant condition, as in typical Fabry-Perot resonators, resulting in a sharp dip in reflection, *SI Appendix Figure 10*. Although this resonance offers colors with higher saturation, the pure geometrical nature of the mode makes it highly angle-dependent, thus limiting greatly its practical applications, offering further proof of the fundamental advantage of the near-field coupled gap-plasmon engineered inside this novel self-assembled ultra-thin structure which is exploited here.

**A Versatile Platform for Structural Coloration.** Growing the structure with conventional evaporation techniques at low temperatures permits the use of a wide variety of substrates. To prove the versatility of the proposed plasmonic self-assembled structure we produced multicolor butterflies by growing several stacks on wing-shaped polyethylene terephthalate (PET) templates, Figure 4a. The polarization and angle-insensitiveness of these color structures readily shows their superiority over many other reported structural approaches. On the one hand, the polarization independency arises from the isotropic character of the disordered self-assembled layer, where nanoislands show no predominant direction or orientation of growth. In Figure 4b we show how, as expected, when photographed with unpolarized, and two orthogonal linearly polarized states, the butterfly assembly shows no appreciable color difference. This particular feature of the multilayer structure is highly important for integration in devices that rely on the use of polarized light such as liquid crystal displays. On the other hand, the subwavelength character of the cavity makes the structure pretty color insensitive to the angle of incidence. Photographies of the blue artistic butterfly at three different combinations of zenith and azimuth angles show clearly that the color is retained regardless of the angle of incidence, Figure 4c. Indeed, upon further study, *SI Appendix F*, we observe the structures retaining their color for angles as large as 60°.

The adaptability to different substrates of this unique large-area, self-assembling based fabrication method paves the path towards the integration of the stack in elastic platforms without the loss of color quality. We grow three samples with 5, 8 and 12 nm nanoparticles' layers, and fixed 10 nm aluminum oxide spacer, on top of aluminum coated polyethylene terephthalate (PET) strips. These three configurations, corresponding to the three primaries in the CYM color mode, can be seen in Figure 4d. Although vivid and brilliant, the specular coloration observed in flat substrates is inconvenient in many applications. For such cases, corrugated substrates can be used to produce diffuse coloration mode. In the diffuse coloration mode, careful texturing of the substrate can control the degree of dispersion of light reflected. We produce diffuse coloration by growing the nanostack on sandblasted PET strips, Figure 4e. In contrast to flat substrates, that result in specular coloration mode, the use of microtextured substrates result in surfaces that homogenously diffuse the light without inconvenient light streaks of specular reflection, while retaining angle and polarization insensitiveness.

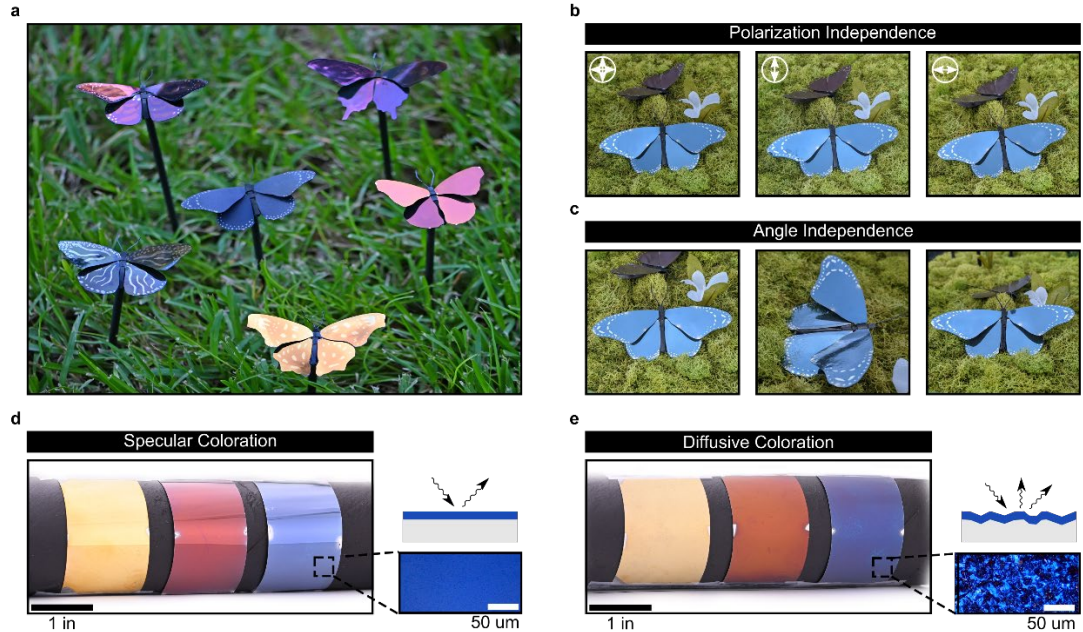

**Figure 4 | Dual Color Mode, Polarization Independence, and Angle-Insensitiveness. a,** Butterfly garden with an assorted collection of different butterfly wings and colors. **b**, An artistic butterfly model coated with structural blue retains its color when photographed with unpolarized –left-, and two orthogonal linearly polarized states –center and right-. **c**, The butterfly color is also angle-insensitive, as shown for three different combinations of azimuth and zenith angles. **d-e**, The versatility of the self-assembly fabrication process permits the use of a wide array of substrates. Flat and sandblasted PET strips are used as flexible substrates to form the three primaries in both **d,** specular, and **e**, diffuse coloration mode.

**Expanding the Color Gamut with Mixing of Structures.** Changes in the geometrical parameters can be introduced to tailor the color response of the structure. Often, however, the production of a larger color palette is difficult, due to the limitation of primary colors. Guided by the principle of conventional color mixing where multiple pigments are mixed to produce secondary colors, we demonstrated in a unique way production of new colors by the "mixing of structures" without needing new materials. Growing side-by-side patches covered with 5 and 10 nm mass thickness nanoislands, we controlled the final color appearance by careful selection of the ratio of the area covered by each one of the particles' configurations. We define the control parameter $\alpha$ as the ratio of the area covered by the 10 nm equivalent nanoislands to the total area covered by both configurations, Figure 5a. Using a lithographic mask we define subpixels of 100 μm length and variable 0 to 100 μm width, in steps of 10 μm, to be covered by 10 nm nanoislands. The rest of the area is then covered by the 5 nm mass thickness nanoislands producing samples with $\alpha$ values ranging from 0 to 100. In this manner we fabricated three samples for cavity length values of 10, 15, and 20 nm, Figure 5b. The pixel geometry is purposely selected to be in a chess-board arrangement with pixels smaller than 100 μm to reduce chromatic aliasing and produce smooth colored surfaces to the naked eye. Microscopy insets for selected samples can be seen in Figure 5c. This side-by-side mixing mechanism can be explained by a simple additive rule. For given reflection curves, $R_A$ and $R_B$, corresponding to the two color bases A and B with mixing ratio $\alpha$, the total reflection is given by:

$$R_{tot} = (1 - \alpha) \cdot R_A + \alpha \cdot R_B$$

Reflection curves for the mixtures with fixed 10 nm-oxide layer can be seen in Figure 5d, where we observe the transition from one basis to the other. This is even clearer in

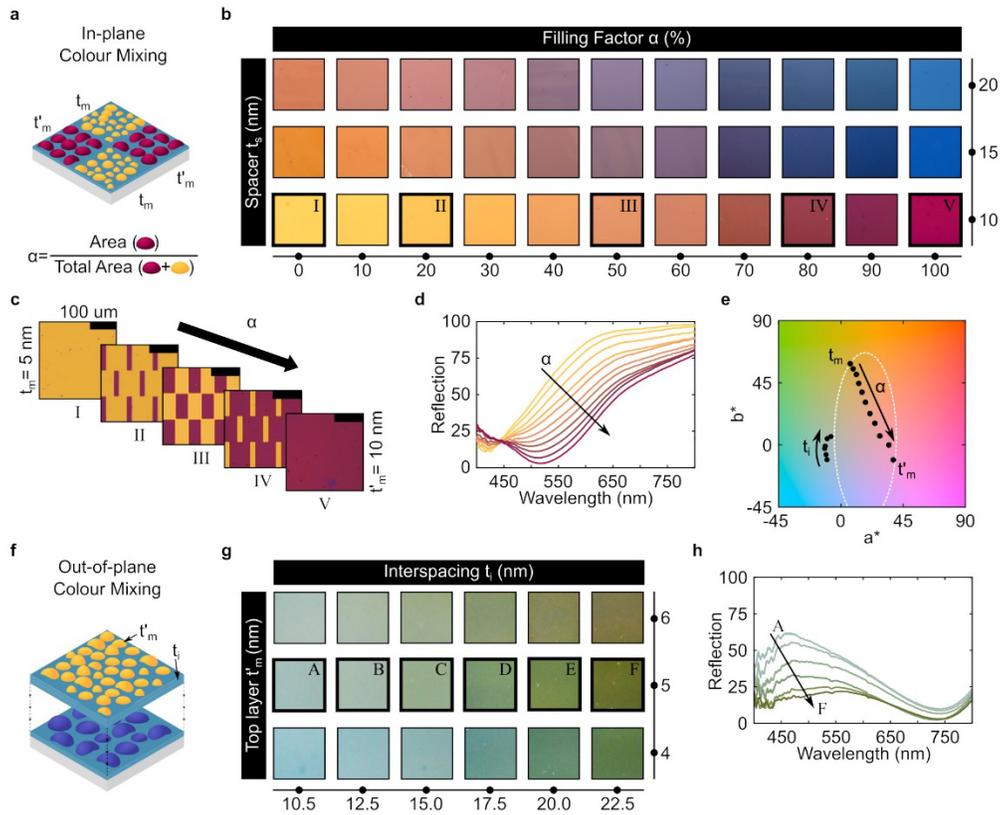

**Figure 5 | Mixing of Structures to Expand the Color Gamut. a**, Controlling the ratio of area covered by two configurations the reflection curve can be defined by a simple additive rule. **b**, Camera pictures of samples with mixing ratios from 0 to 100%, for spacer thicknesses of 10, 15, and 20 nm. **c**, Microscope images for the samples highlighted in b. **d**, As the ratio is increased the reflection curves transition from pure basis A to pure basis B. **e**, CIELAB space for the samples corresponding to spacer thickness of 10 nm in panels b and g. The white dotted line overlay represents the space defined by the color wheel in Figure 3c. **f**, New colors can be generated by multilayer structures. **g**, Green shades inaccessible with a single layer can be generated by stacking two self-assemblies with different interspacing thicknesses. **h,** Tuning of the interspace layer between self-assemblies controls the optical response of the cavity.

Figure 5e, where we have plotted as black dots the L*a*b* coordinates corresponding to the $\alpha$ values from 0 to 100. For context, the colorspace defined by the thickness wheel analyzed in Figure 2c is overlaid as dotted white lines. Conveniently enough, any color contained in the segment defined by the two coordinates corresponding to the bases can be generated by careful selection of in-plane mixing ratios, *SI Appendix G*.

In-plane mixing does expand the color palette by offering a route to generate any color contained in the region defined by the basis employed. However, it does not permit to generate colors outside of its boundaries. Generating green shades would therefore require a green basis. Yet, due to its subtractive nature, the production of green is prohibited for the plasmonic self-assembly. However, this limitation can be broken by growing multilayers of plasmonic nanoparticles, Figure 5f. In this multilayer configuration two extra geometrical parameters are introduced to control the color appearance: the thickness mass of the extra layer and the interspace between the nanoislands films. We produce a wide variety of green shades by growing, on top of a base structure consisting of an aluminum mirror, 10 nm oxide layer, and 10 nm equivalent nanoislands, three top layers corresponding to 4, 5, and 6 nm self-assembled layers with oxide interspaces ranging from 10.5 to 22.5 nm, Figure 5g. The reflection curves for the 5 nm equivalent top layer can be seen in Figure 5h, while the L*a*b* coordinates for these curves are plotted as dots in Figure 5e, where we observe how the out-of-plane mixing scheme does indeed expand the color palette to areas otherwise inaccessible with a single plasmonic layer. Although the levels of reflection in the bilayer structures are low, due to the double absorption of the two-fold plasmonic layer, careful study of all geometrical parameters can help mitigate partially this effect. Indeed, the interspace and top-layer geometrical parameters add to the bottom self-assembly and the spacer layer thickness to offer extra degrees of freedom to expand the available color gamut introducing new colors with different saturation and luminance, *SI Appendix H*. Interestingly, from the industrial point of view, these new layers do not increase excessively the complexity of the manufacturing, as they can be grown in the same chamber as a single process step, with little extra time or cost. The addition of an extra plasmonic film can also be incorporated to the analytical model by adding on top of the single stack two extra layers: the dielectric interspace and an extra effective medium describing the upper plasmonic layer.

**Structural Color Paint.** The fabrication process used to grow the self-assembled structure permits an easy integration in many industrial processes that already use compatible systems. Despite this, the platform is fundamentally limited by the need to grow in-situ structures. This restricts critically the applicability in contexts where non-vacuum-compatible substrates are required or instances where large areas need to be covered, as the limiting factor would always be the particular specification of the evaporation equipment used. To present a realistic alternative to commercial chemical colorants the multilayer structure should ideally be available in a stand-alone platform that can be transferred, after fabrication, to any substrate. On top of a sacrificial layer we evaporate sequentially a double-sided mirror-symmetric stack, where each side comprises an alumina protecting capping layer, a plasmonic self-assembly and an alumina spacer, while the mirror is shared between them. Removal of the sacrificial layer at the end of the fabrication results in tunable self-standing doubly-colored flakes, Figure 6. We chose to grow the structures symmetrically to ensure homogeneous colorization, however, flakes can be grown in asymmetric configurations, each face showing a different color, to render new mixtures similar to the in-plane mixing color scheme explored before. Once the structures are flaked off of the substrate, the flakes can be stored dry in powder form, Figure 6b, or kept in an organic solvent (here we used acetone), Figure 6c. After lifting off, flakes present irregular shapes and sizes, with lateral dimensions 20-150 μm. To increase homogeneity and improve efficiency when coating, a final ultrasonication and filtering step is performed to break flakes to lateral sizes of 10s of μm, *SI Appendix I*.

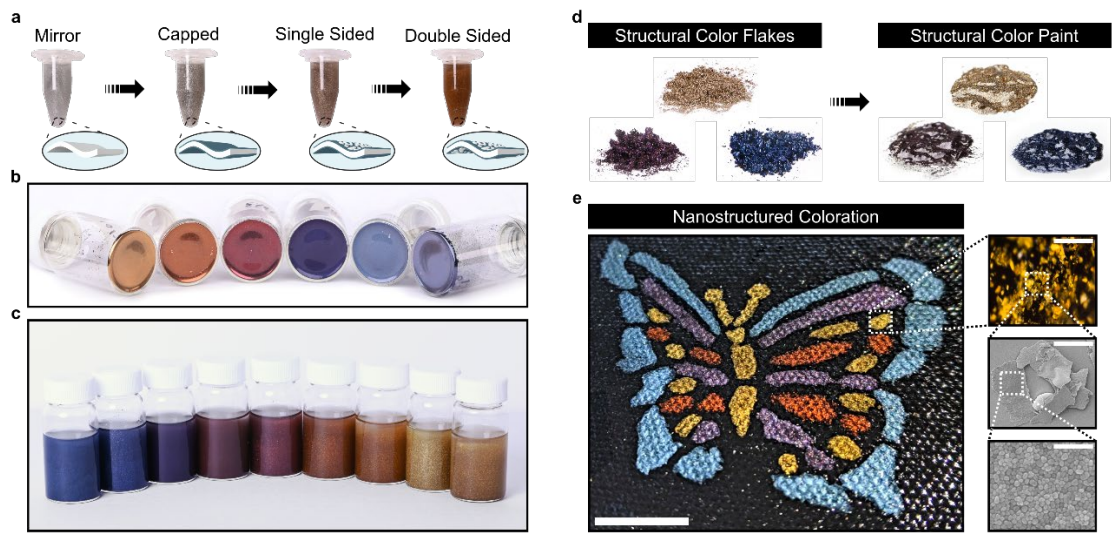

**Figure 6 | Structural Color Paint. a** Sequential growing of a bi-directional stack on a sacrificial layer results on color flakes. **b** and **c**, Color flakes can be stored dry or dispersed in a solution. **d**, A paint can be produced by mixing the flakes with a drying oil. This simple formulation, where the flakes are the pigments and the oil the binder, can be adapted to impart the nanostructured coloration to any surface. **e**, Photography of a multicolor artistic butterfly on a black canvas painted with a set of linseed oil-based plasmonic paints demonstrating the commercial feasibility of the platform. Insets correspond to a microscope image –top- and SEM micrographs –bottom-. Scale bar for the butterfly is 1 inch, whereas for the insets, from top-left to bottom-right, scale bars corresponds to 1 mm, 100 μm, 75 μm, and 100 nm, respectively.

To demonstrate the commercial potential of this platform for inorganic metallic pigmentation, we formulated a paint by mixing the structural color flakes with a drying oil (Linseed oil, Gamblin), Figure 6a. The mixture presents the simplest form of a paint, where the flakes are the pigment and the oil is the binder that permits transferring to the target substrate. This mixture can be used then to coat surfaces in applications otherwise incompatible with vacuum systems. In Figure 6b we show such an example, where we painted an artistic multicolor butterfly on a black canvas. Although different target surfaces would require more careful selection of the binder, and possibly the use of other chemical additives in the paint formulation, the structural color paint demonstrated here can be easily adapted. Indeed, provided that non-corrosive chemicals are used, the flakes self-assemble structure is a universal platform independent of the particular paint components employed. Finally, from a commercial point of view, two additional features make this structural color paint a very promising candidate for industrial production. First, in contrast to chemical coloration schemes that use toxic and contaminant components, the fabricated flakes avoid detrimental environmental impacts by employing only nontoxic materials such as aluminum and its oxide, and a biodegradable water-soluble polymer as a sacrificial layer (see Methods). And second, the structural color paint offers 100% reflection with only a single ultra-thin layer of thickness (100 – 150 nm) pigment of extremely low surface density. [22]

## Conclusion

While structural coloration presents a promising opportunity to substitute chemical colorants with purer and non-toxic options, commercial production of such structural color poses a challenge due to the anisotropic optical response that results in undesired effects such as dichroism or iridescence coupled with tedious fabrication processes. In this work we have demonstrated a color nanostructure that overcomes these challenges offering a real-world opportunity for industrial production.

Hybridizing the plasmonic response of a metallic self-assembly with an ultrathin optical cavity we have demonstrated a large CYM palette that can be produced by simply changing the geometrical parameters of the structure. Furthermore, we studied mechanisms for expanding the available color space through multilayers and in-plane addition for cost-effective color mixing to expand the color gamut further. While the isotropic character of the nanoislands' layer ensures the polarization independence, the angle insensitiveness, otherwise impossible in conventional Fabry-Perot resonators, is achieved by exploiting the non-trivial phase discontinuities in the ultrathin cavity, thus avoiding path length effects for steep angles close to 70°.

The subwavelength plasmonic cavity is fabricated with a low-temperature process in an ebeam evaporator. The versatility of the process permits the use of many different substrates, including flexible platforms required in wearable electronics and roll-to-roll manufactures, and takes on the scattering properties of the target surface to produce both diffuse and specular coloration mode. Additionally, being a self-assembly process, the color consistency is ensured for large areas. We observed that color purity of our structure is dependent on the size and shape distribution of the nanoislands. We note that although the use of pre-seeding techniques, higher-temperatures, or alternative materials would improve control on the assembly morphology it would impose a scale and cost toll on the manufacturing process.

To demonstrate the commercial capabilities of our platform for inorganic metallic pigments, we have fabricated self-standing bi-directional color flakes by evaporating on top of a water-soluble sacrificial layer a double-sided stack. This ultralight pigment, that offers full coloration with a single layer of flakes, can be then mixed with a binder matrix to formulate a structural color paint that can be used to coat, after fabrication, any surface. This approach presents an ultralight, multi-color, large-scale, low-cost, and environmental-friendly platform for imparting nanostructured coloration to any surface. With an unbeatable surface density of 0.4 g/m$^2$, hundreds of times lighter than commercially available paints, thus paving the way towards industrial production and real-world application.

___

**Methods**

**Self-Assembled Structural Color Fabrication.** The optically thick back-mirror is produced by evaporating 100 nm of aluminum on a Thermionics e-beam evaporator. Pressure at the beginning of the evaporation was ~1x10$^{-6}$ Torr, and evaporation rate was kept at 0.1 nm/s. A spacer layer of aluminum oxide was then grown by atomic layer deposition (Savannah 200, Cambridge Nanotech) by pulsing trimetylaluminum and water at 100 °C. The aluminum nanoparticles were then produced on an UHV AJA electron beam evaporator. In agreement with previous studies[17,31,32], we find that three parameters play a critical role in the geometry of the self-assembled monolayer: the temperature of the substrate, the pressure in the chamber, and the rate of growth. We chose these parameters compromising between the desired higher saturation of colors and the lower requirements for fabrication that ensures the versatility of the proposed architecture and the viability of a transition to industrial-scale production. Thus, we chose to keep the temperature of the substrates at 100 °C, resulting in high color saturation while being below the melting point of many polymers, essential for flexible substrate applications and lift-off during flake preparation. For reproducibility and color vividness, nanoparticles' growth was carried on at pressures below 5x10$^{-8}$ Torr, while growth rates were kept constant about 0.1 A/s, both readily available in conventional UHV evaporators. It should be note that although an ALD system was used for convenience for the hard substrates, the entire fabrication process could be carried on in a single UHV ebeam evaporator. To prove this promising ease of integration, critical for industrial chain systems, the oxide layer in the pigment flakes production was grown at room temperature on the Thermionics ebeam evaporator system. Besides a slight change in color shade, attributed to the change in refractive index of the oxide layer, no fundamental quality change was observed between the layers grown by the two systems.

**Finite Difference Time Domain Modeling.** The reflection spectra and electric field distribution of the simulated samples were calculated using the experimental geometrical parameters extracted from the SEM analysis of the samples, with a commercial FDTD software package (Lumerical FDTD, Lumerical Solutions Inc.). The relative permittivities of aluminum and aluminum oxide are taken from literature[33]. To build the weight-averaged semi-analytical model of *SI Appendix Figure 2* we simulated a single particle with periodic conditions. The values of the unit cell size were chosen to make the area covered by the aluminum island equivalent to those obtained from the SEM analysis for the different samples (55%, 60% and 70% for the 4, 8 and 12 nm respectively). 50 simulations with radii values within 4 standard deviations of the mean value were then performed and the weight-averaged reflectance calculated as:

$$\bar{R} = \sum_i R_i \times w_i$$

where $w_i$ and $R_i$ represent the Gaussian weight and simulated reflection for the particle $i$. For the analysis of the effect of the disorder as shown in *SI Appendix Figure 3*, we simulated an array of 49 particles placed in periodic arrangement, disordered arrangement, and disordered arrangement of different radii, introducing the disorder parameter following the method presented by Zhang et al.[34].

**Color Gamut Evaluations.** To find the L*a*b* coordinates of the fabricated samples we first obtained the XYZ tristimulus values integrating over the visible spectrum according to:

$$X = \frac{1}{N} \int \bar{x}(\lambda) R(\lambda) I(\lambda) d\lambda$$

$$Y = \frac{1}{N} \int \bar{x}(\lambda) R(\lambda) I(\lambda) d\lambda$$

$$Z = \frac{1}{N} \int \bar{x}(\lambda) R(\lambda) I(\lambda) d\lambda$$

$$N = \int \bar{y}(\lambda) I(\lambda) d\lambda$$

where $R(\lambda)$ is the measured reflectance; $\bar{x}(\lambda)$, $\bar{y}(\lambda)$, and $\bar{z}(\lambda)$ are the color matching functions; $I(\lambda)$ is the reference illuminant. From the tristimulus values CIEXYZ the CIELAB coordinates can be calculated from:

$$L^* = 116 f_y - 16$$

$$a^* = 500 (f_x - f_y)$$

$$b^* = 200 (f_y - f_z)$$

where, being $t_x = \frac{X}{X_n}$, $t_y = \frac{Y}{Y_n}$, or $t_z = \frac{Z}{Z_n}$; and $X_N = 0.9642$, $Y_N = 1.0000$, and $Z_N = 0.8251$ the D50 white point coordinates:

$$f_i = \begin{cases} \sqrt[3]{t_i} & \text{if } t > \left(\frac{6}{29}\right)^3 \\ \frac{841\, t_i}{108} + \frac{4}{29} & \text{otherwise} \end{cases}$$

As the reference illuminant we chose D50, used in the graphic arts industry for color proofing as per ISO 3664:2009. Finally, to keep consistency, all colormaps presented in the figures in the main manuscript and the *SI Appendix* correspond to a horizontal slice with $L^* = 75$.

**Measurements and Images.** Reflection measurements were taken at normal incidence with unpolarized light using a 4x, 0.07 numerical aperture objective and a fiber coupled spectrometer (HR 2000+, Ocean Optics). An aluminum mirror was used as a reference. Angular measurements were taken with an integrating sphere (RTC-060-SF, Labsphere) connected to the spectrometer. To ensure consistency on illumination the samples were photographed with flash-light at fixed intensity. Photographies for the butterfly models were taken under sun illumination with a linear polarizer attached to the objective.

## Acknowledgments
This work at University of Central Florida was supported by National Science Foundation Grant #ECCS-1920840. P.C.A. acknowledges the support from the UCF Preeminent Postdoctoral Fellowship Program (P3).

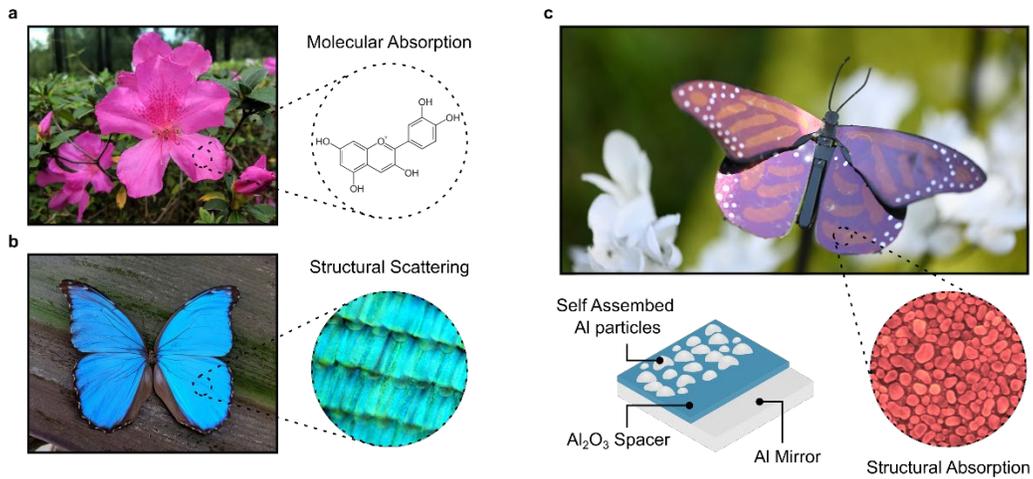

**Figure 1 | Structural Absorption for Color Generation**. **a**, Many chemical substances produce color by selectively absorbing frequencies matching their molecular electronic transitions. Pink color in Formosa azaleas is due to the absorption of cyaniding molecules. **b**, An example of structural coloration is found in the Peruvian Morpho didius. Lamellae nanostructures found in its wings scatter the blue components of incident light generating its characteristic metallic blue. **c**, A subwavelength plasmonic cavity formed by a self-assembly of metallic nanoislands on top of an oxide-coated mirror, generates color by selectively absorbing certain wavelengths and strongly back-reflecting other.

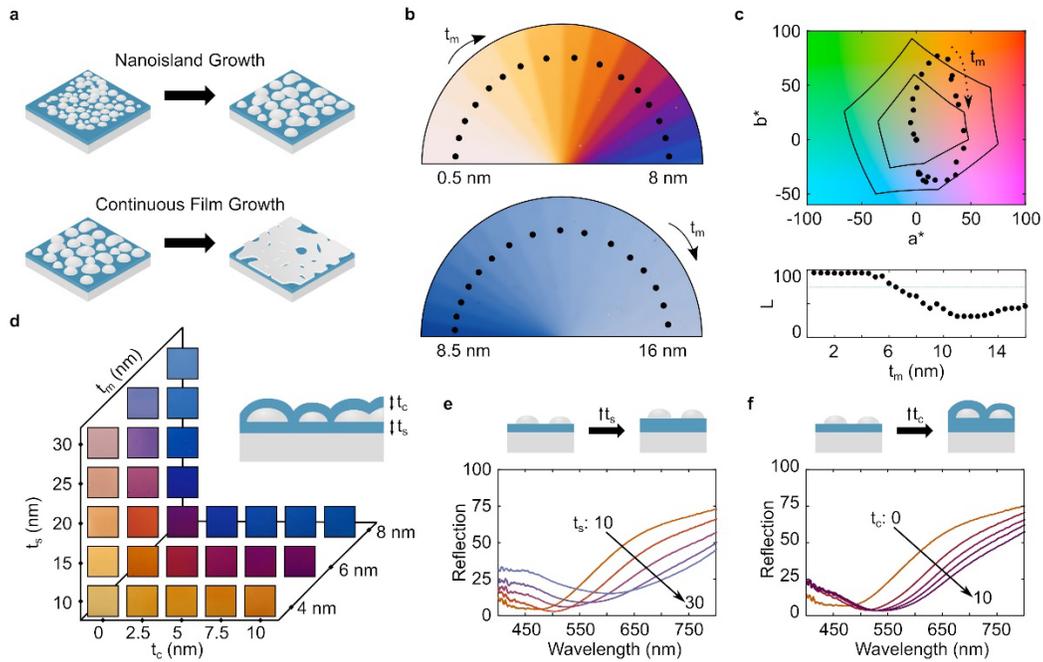

**Figure 2 | Color Space and Quality of the Plasmonic Cavity. a**, In the Volmer-Weber growth mode, the size of the nanoislands can be controlled by tuning the amount of aluminum evaporated –top-. If the process is carried on for long enough semicontinuos films are formed that disable the plasmonic resonances and thus the color –bottom-. **b**, Color polar gradient for thicknesses from 0.5 to 16 nm, for fixed 10 nm spacer. **c**, CIELAB coordinates for the points in the color wheel compared to ISO DIS 15339-2 cold-set newsprint and coated premium paper standards (inner and outer hexagon). **d**, Tuning of the spacer and capping layer thicknesses expands the available color space. **e**, **f**, Show the red-shift of the absorption resonance as the spacer and capping layer thicknesses are increased.

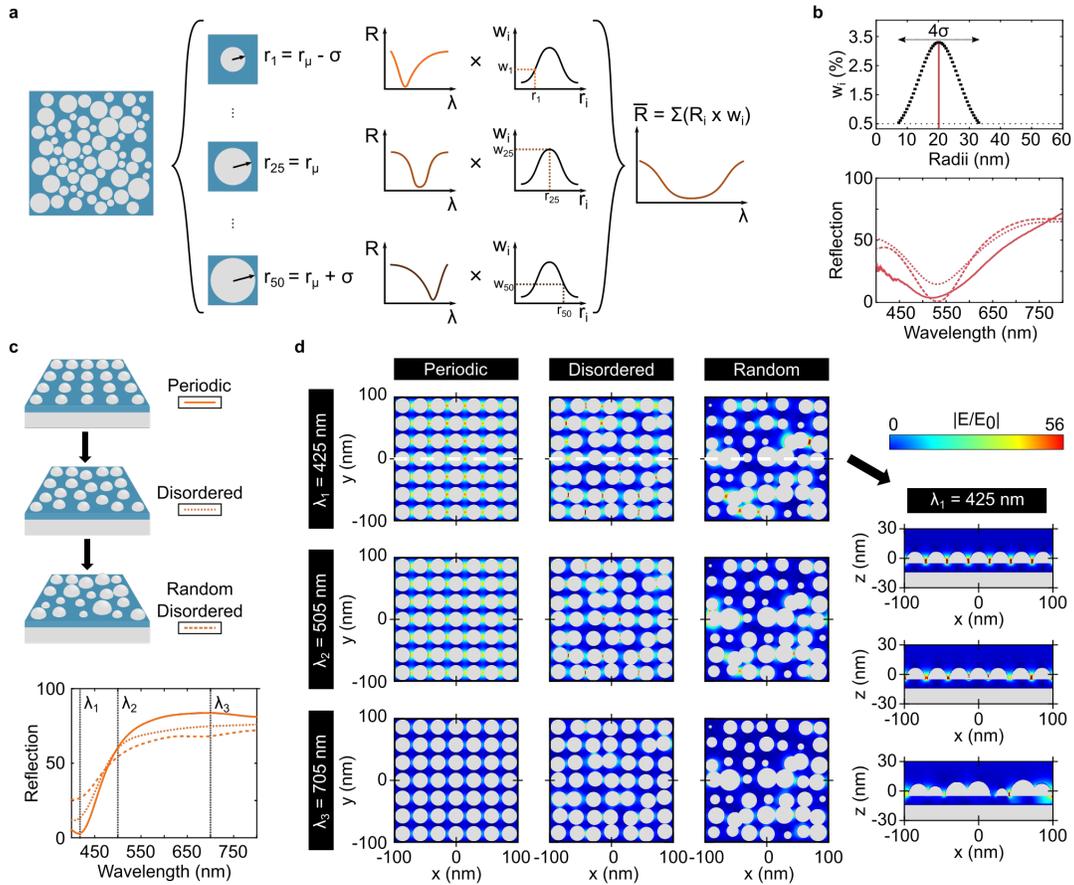

**Figure 7 | Morphology Effect on Optical Response. a**, The inhomogeneous broadening of the optical resonance can be accounted for by introducing size and spatial variability. We simulate the broadening by averaging the reflection curves of 50 particles with radii within 4 standard deviation of the mean value obtained from SEM analysis. **b**, Statistical radii distribution -top- and reflection curves -bottom- for experimental (solid), FDTD mean value (dashed), and weight averaged (dotted). **c**, Reflection curves for FDTD simulations corresponding to 7x7 hemispherical particles with equal size in periodic and disorder arrangement, and random size and disordered arrangement, equivalent to the 5 nm self-assembly. **d**, Electric profiles in three different spectral positions as labeled in panel c.

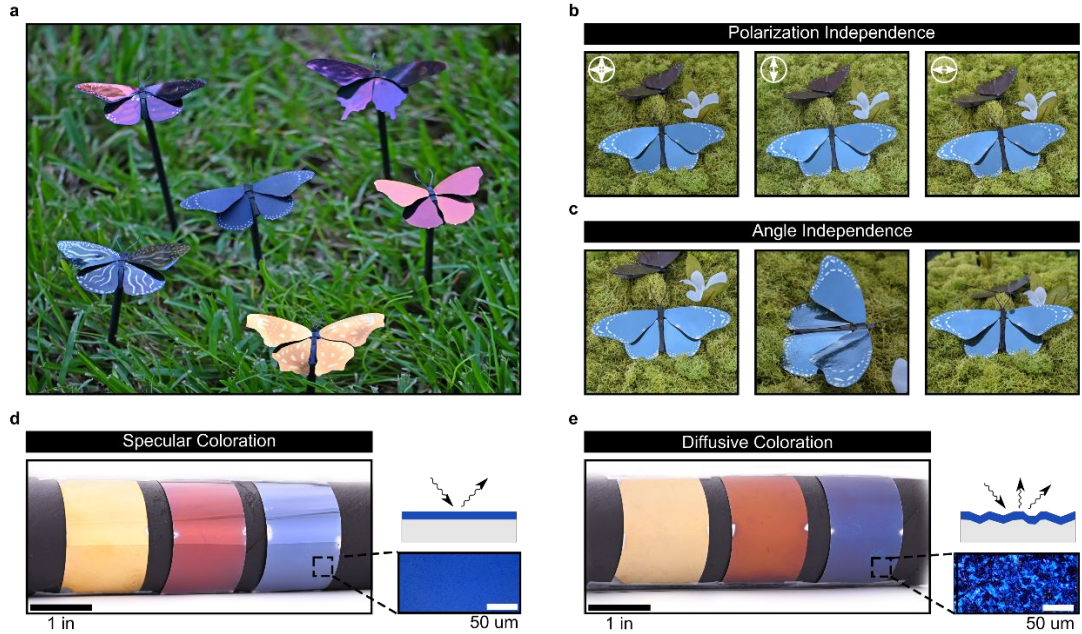

**Figure 4 | Dual Color Mode, Polarization Independence, and Angle-Insensitiveness. a,** Butterfly garden with an assorted collection of different butterfly wings and colors. **b**, An artistic butterfly model coated with structural blue retains its color when photographed with unpolarized –left-, and two orthogonal linearly polarized states –center and right-. **c**, The butterfly color is also angle-insensitive, as shown for three different combinations of azimuth and zenith angles. **d-e**, The versatility of the self-assembly fabrication process permits the use of a wide array of substrates. Flat and sandblasted PET strips are used as flexible substrates to form the three primaries in both **d,** specular, and **e**, diffuse coloration mode.

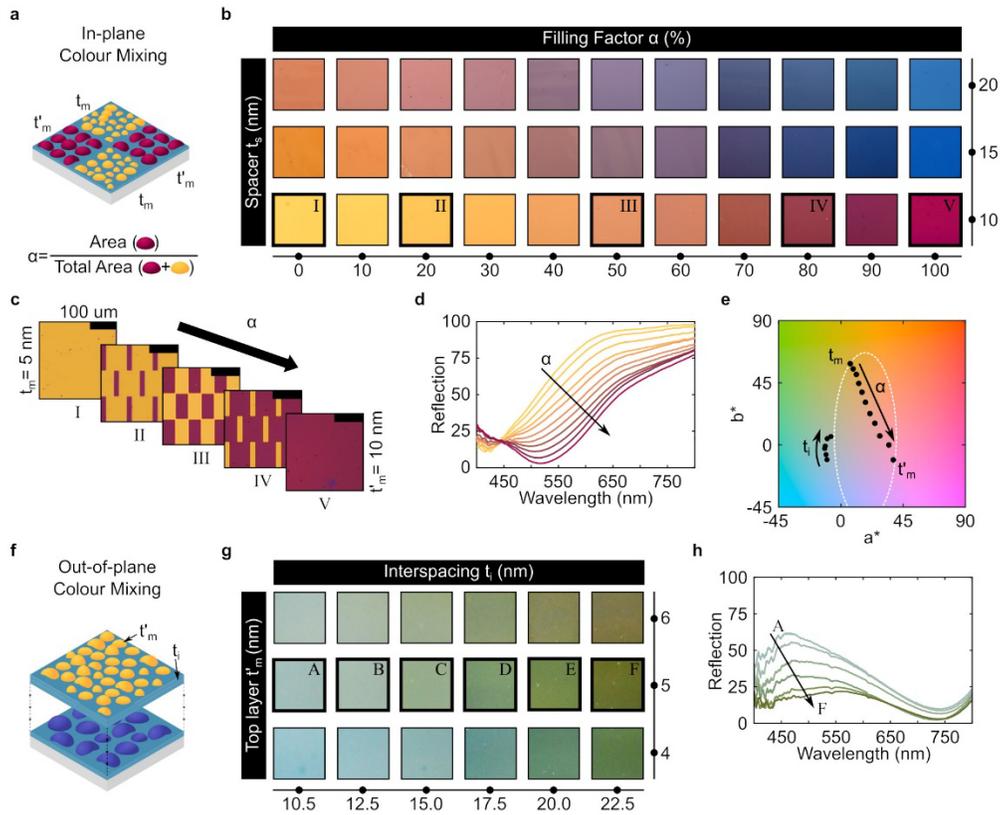

**Figure 5 | Mixing Schemes for Expanding the Color Palette. a**, Controlling the ratio of area covered by two configurations the reflection curve can be defined by a simple additive rule. **b**, Camera pictures of samples with mixing ratios from 0 to 100%, for spacer thicknesses of 10, 15, and 20 nm. **c**, Microscope images for the samples highlighted in b. **d**, As the ratio is increased the reflection curves transition from pure basis A to pure basis B. **e**, CIELAB space for the samples corresponding to spacer thickness of 10 nm in panels b and g. The white dotted line overlay represents the space defined by the color wheel in Figure 3c. **f**, New colors can be generated by multilayer structures. **g**, Green shades inaccessible with a single layer can be generated by stacking two self-assemblies with different interspacing thicknesses. **h,** Tuning of the interspace layer between self-assemblies controls the optical response of the cavity.

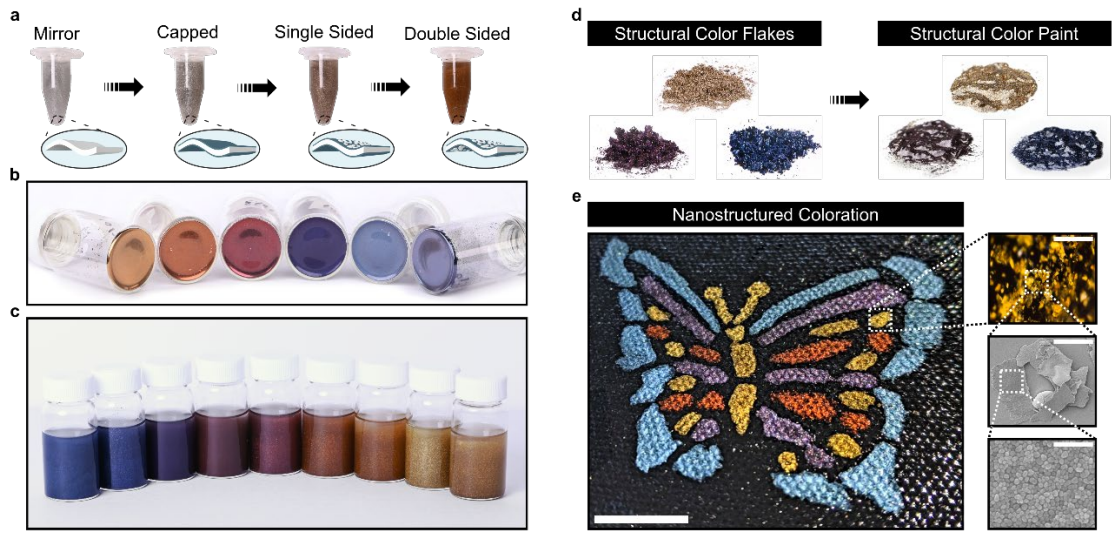

**Figure 6 | Structural Color Paint. a** Sequential growing of a bi-directional stack on a sacrificial layer results on color flakes. **b** and **c**, Color flakes can be stored dry or dispersed in a solution. **d**, A paint can be produced by mixing the flakes with a drying oil. This simple formulation, where the flakes are the pigments and the oil the binder, can be adapted to impart the nanostructured coloration to any surface. **e**, Photography of a multicolor artistic butterfly on a black canvas painted with a set of linseed oil-based plasmonic paints demonstrating the commercial feasibility of the platform. Insets correspond to a microscope image –top- and SEM micrographs –bottom-. Scale bar for the butterfly is 1 inch, whereas for the insets, from top-left to bottom-right, scale bars corresponds to 1 mm, 100 μm, 75 μm, and 100 nm, respectively.